\documentclass[12pt]{article}
\usepackage{rotating}
\usepackage{epsfig,amssymb}
\input{epsf}

\def\laq{\raise 0.4 ex \hbox{$<$}\kern -0.8 em\lower 0.62 ex\hbox{$\sim$}}
\def\gaq{\raise 0.4 ex \hbox{$>$}\kern -0.7 em\lower 0.62 ex\hbox{$\sim$}}

\def\beq{\begin{equation}}
\def\eeq{\end{equation}}
\def\beqa{\begin{eqnarray}}
\def\eeqa{\end{eqnarray}}

 \def\frac#1#2{{\textstyle{{#1}\over {#2}}}}
 \def\lsim{\mathrel{\rlap{\lower4pt\hbox{\hskip1pt$\sim$}}
    \raise1pt\hbox{$<$}}} \def\gsim{\mathrel{\rlap{\lower4pt\hbox{\hskip1pt$\sim$}}
    \raise1pt\hbox{$>$}}}
\def\sqr#1#2{{\vcenter{\vbox{\hrule height.#2pt
         \hbox{\vrule width.#2pt height#1pt \kern#1pt
         \vrule width.#2pt}
         \hrule height.#2pt}}}}

\def\gappeq{\mathrel{\rlap {\raise.5ex\hbox{$>$}} {\lower.5ex\hbox{$\sim$}}}}

\def\lappeq{\mathrel{\rlap{\raise.5ex\hbox{$<$}}
{\lower.5ex\hbox{$\sim$}}}}

\begin{document}
\pagestyle{plain}

\begin{flushright}
\end{flushright}
\vspace{5mm}

\begin{center}

{\Large\bf The Physics of Time: Current Knowledge and Unanswered Challenges\footnote{Based on talk delivered at the 13th Bial Symposium, ``The mystery of time", Behind and Beyond the Brain, 6-9 April 2022, Porto, Portugal.}}

\vspace*{0.2cm}

Orfeu Bertolami$^{1,2}$  \\
\vspace*{0.5cm}
{$^1$Departamento de F\'\i sica e Astronomia, Faculdade de Ci\^encias, Universidade do Porto\\
Rua do Campo Alegre s/n, 4169-007 Porto, Portugal}\\
\vspace*{0.2cm}
{$^2$Centro de F\'\i sica das Universidades do Minho e do Porto, Faculdade de Ci\^encias, Universidade do Porto\\
Rua do Campo Alegre s/n, 4169-007 Porto, Portugal}\\

\vspace*{0.2cm}
\end{center}

\begin{abstract}

\noindent
In contemporary physics space and time are intertwined entities so that kinematical and dynamical quantities are expressed in the four-dimensional space-time. 
This formulation seems to contradict our every-day experience and perception according to which space and time are distinct entities. In this brief report 
we shall discuss these apparently antagonist views and analyse the underlying physical property of time, namely that it evolves from the past to the present, from the present to the future.

\end{abstract}


{\noindent
Time passed, turning everything to ice. \\
Under the ice, the future stirred. \\
If you fell into it, you died. \\

\noindent 
It was a time \\
of waiting, of suspended action.

\noindent 
I lived in the present, which was \\
that part of the future you could see.\\
The past floated above my head,\\
like the sun and moon, visible but never reachable.

\noindent 
It was a time \\
governed by contradictions, as in \\
I felt nothing and\\
I was afraid.}

\noindent
in Averno (2006)

\noindent 
Louise Gl\"uck

\section{Introduction}

According to the Big-Bang theory, the origin of the Universe is, in fact, the process of creation of matter, space and time. Despite its modernity and sophistication, 
the science that describes this extraordinary event, shares the idea of a primordial beginning with the very first manifestations of 
rational thinking of distinct civilisations, the so-called myths of creation (see, for instance \cite{Farrell2005,Bertolami2006} 
for extensive discussions). The earliest forms of philosophical thinking, have invariably regarded space and time as essential features of reality, even though quite distinct ones. 

Indeed, at first glance, space and time seem to be quite different. Space can be freely experienced as one can move in any direction without restriction. Time however, has a 
well defined direction. Past and future are clearly distinct as our actions can affect only the latter. We have memory, but not precognition. Matter, organic or otherwise, eventually decay 
ceaselessly rather than getting itself organised. As far as we can detect through our most powerful microscopes, the particle accelerators, the fabric of space-time is made of 
at least three distinct spatial dimensions and only one time dimension\footnote{Notice that if the number of time dimensions were greater than one, one would encounter all types of 
complications as, on quite general grounds, the Partial Differential Equations that describe the physical phenomena would be ultra-hyperbolic, leading to unpredictability, or to 
backward causality (see e.g. Ref. \cite{Bertolami2007} for a discussion).}. 

Actually, in physics, time allows for describing the evolution and transformation of systems and processes in a very convenient way, being for this reason an ``invisible" or ``hidden" dimension. 
Despite of that, time is similar to space, as both have always been thought as the arena of all manifestations of Nature. Everything lies in space, as well as in time, 
and the intrinsic and fundamental relationships between the most basic elements of all physical entities could be decomposed into points, straight lines and geometrical 
figures in two or three dimensions 
and whose properties were remarkably systematised by Euclid (ca. 330 - 275 B.C.) geometry. These relationships would in turn reveal the intrinsic properties of space itself. 
These geometrical properties are shared by time with the unification of space and time 
proposed by Hermann Minkowski \cite{Minkowski} in 1908, inspired by Einstein's Theory of Special Relativity 
(for a discussion see, for instance, \cite{Bertolami2008a,Bertolami_Lobo}). 

On the other hand, by any account, time is a mysterious ingredient of the Universe and has been the subject 
of continuous philosophical debate over the centuries (we refer the interest reader to the brief discussion in Ref. \cite{Bertolami2008a}).
Time is intuitively related to change, and it is
subjectively perceived as a permeating entity that flows. This
view can be traced back as far as to Aristotle (384 BC - 322 BC), who regarded 
``time as the measure of change''. Throughout history, one encounters a wide range of
reflections and considerations about the nature of time. For many civilisations, cycles in Nature, were an
evidence of the circular and repetitive pattern of time. It was only in
the 17th century that Francis Bacon (1561 - 1626) posed 
the concept of linear time, and through the influence
of Newton (1643 - 1727), Barrow (1630 - 1677), Leibniz (1646 - 1716), 
Locke (1632 - 1704) and Kant (1724 - 1804) amongst others, about the
19th century, the idea of a linear time evolution was the dominant one in science and philosophy 
(for discussions see, for instance, 
\cite{Bertolami2006,Bertolami2008a,Russell,Prigogine1,Montheron,Ellis,Tipler80,Coveney}). 
However, despite all evidence and arguments, for some physicists, Einstein being amongst the finest, the irreversible flow of time is 
no more than a ``persistent illusion".

In what concerns its ontological nature, it is still a matter of debate whether
time is a real fundamental quantity or, instead, a composite suitable parameter, like {\it temperature} to 
describe the physical laws (see, for instance, Ref. \cite{Seiberg} for a discussion in the context of string theory) and to frame, 
in unequivocal terms, causation. Causation being, of course, a
crucial feature not only in
physics, but according to some philosophers, 
like David Hume (1711 - 1776), a fundamental element for the  
human understanding of reality.
In fact, a relevant connection to physics arises here as the decomposition of human perception down to the 
physiology of nervous tissues, down to its chemistry and then down to the causal character of the physical laws, 
renders Hume's proposition quite plausible. 

Back to physics, one could state that
reflections about the nature of time were polarised by Newton's concept of absolute
time, who assumed that time flowed at the same rate for all
observers in the Universe \cite{Newton}. For Newton, time and
space were an infinitely large vessel, containing events, and
existing independently of the latter. This picture was completely changed in 1905 by 
Albert Einstein (1879 - 1955), through the
formulation of the Special Theory of Relativity, in the context of which it was shown, in 
particular, that time flows at different rates for different
observers. Three years later, Hermann Minkowski (1879 - 1909) \cite{Minkowski}
suggested the unification of the time and space, giving
rise to the notion of a fundamental four-dimensional entity, {\it space-time} mentioned above. The Relativity revolution was concluded in 
1915 when Einstein put forward the Theory of General Relativity where 
it was shown that the {\it flat} spacetime of Special Relativity is curved by energy-matter. Since then the discovery 
of new forms of energy-matter in physics has given rise to new space-time geometries 
(see, for instance, \cite{Bertolami2007} and references therein). 

The merging of space and time into the space-time concept is particularly relevant and 
useful for the physical description of any phenomena in Nature independently of the frame of reference used to
parametrize change-evolution-transformation. It is interesting that this picture that is so attractive for physicists was strongly 
rejected, for instance, by the philosopher Henri Bergson (1859 - 1941). According to Bergson, intelligence and intellect can only have a clear idea of discontinuity and immobility, 
being thus unable to fully grasp life and to think about evolution. For Bergson, intellect tends to represent becoming as a series of states. Geometry and logic, 
as products of intelligence, must be verified against common sense. He believed 
that intelligence and the origin of material bodies are correlative and have evolved by reciprocal adaptation. For him,  
intellect is the power to distinguish different things. However, in reality, there are no separate solid things, 
only a continuous stream of becomings. 
Becoming leads to life when it is an ascendant movement, and to matter when it is a descendant one. 

For Bergson, space and time are profoundly dissimilar. The intellect is associated with space, whilst instinct and intuition are connected with time. 
Space, a feature of matter, arises from a dissection of the flux which is really illusory, although useful in practice. Time, 
on the contrary, is the essential element of life. Mathematical time, according to Bergson, is actually a form of space, 
but the time, as the essence of life, is what he refers to as duration. Duration merges past and present into an organic whole, 
where there is mutual entanglement and succession with distinction.  To an extensive account of the differences between 
Bergson's ideas and the ones pertaining physics and the debate with Einstein in 1922 in Paris, 
the reader is referred to the superb account by Jimena Canales \cite{Canales} and to her talk at this Symposium. 

In what follows we shall further delve in the two main issues related to time. The first one refers to its flow, which concerns 
the so-called arrows of time, associated to the fundamental and ubiquitous Second Law of Thermodynamics. The second one on 
whether the fabric of space-time admits reversible motion in time through closed time-like curves (CTCs) or the so-called ``time machine solutions" of the field equations of General Relativity.

This brief account about the main physical features of time has the following outline: In Section
\ref{Sec:II}, we discuss the issue of directionality of the time flow. In Section \ref{Sec:III}, we address the view of the time in Relativity,  the CTCs 
and the ensued putative violations of causality. Finally in Section \ref{Sec:IV}, we 
briefly discuss some particular issues such as the idea of {\it time crystals}, of a {\it Block Universe},  where all moments in past time still exist and not only the present, and the 
{\it specious present asymmetry}. Of course, this list is quite short and many other issues could be addressed, such as for instance, the putative connection among the various arrows of time 
and the problem of time in quantum gravity. They certainly deserve a discussion, however, given their technicality, we refer the interested reader 
to Ref. \cite{Bertolami_Lobo}. Finally in Section 5 we 
present our closing remarks.

\section{Arrows of time}\label{Sec:II}

A fundamental feature of the nature of time concerns its flow. Modern physics 
describes phenomena through dynamical laws, from which the time evolution 
is determined after specifying a suitable set of initial conditions. 
The fundamental dynamical equations of
classical and quantum physics are symmetric under a time
reversal, i.e., mathematically, one could instead specify the
final conditions and evolve the physical system back in time. 
The resulting would be indistinguishable from the one obtained by 
the usual procedure. 
However, in macroscopical phenomena, which are described by 
thermodynamics, as well as some instances in general
relativity and quantum mechanics (at the measurement process) , the evolution of the systems
is essentially asymmetric in time. This distinguishes past from future. 
In the context of the Second Law
of Thermodynamics, which states that in an isolated system, 
{\it entropy}, a measure of microscopic disorder, is a strictly increasing function of state providing itself a
direction for the flow of time. It is an interesting conjecture the one which
the Second Law of Thermodynamics and the thermodynamical arrow of
time follow from the initial conditions of the Universe, meaning 
that the flow of time is set by 
the expansion of the Universe and it would revert in case of contraction.

In fact, an intense discussion on the physics of macroscopic phenomena took place in the second half of XIX century, 
when Maxwell (1831 - 1879) and Boltzmann (1844 - 1906) became aware of the tension between the linear time evolution 
and the reversibility of the fundamental equations of motion and later on with 
the eternal recurrence of motion in the phase space (the composed space of positions and velocities of the particles), shown by Poincar\'e (1854 - 1912) in 1890. 

Before that, in the 1870s, Boltzmann understood that getting an arrow of time out the mechanics of
atoms was impossible without using averaging arguments. He eventually obtained in 1872 a
time-asymmetric evolution equation, now referred to as 
Boltzmann's equation, whose solution was a single-particle time dependent 
distribution function of a set of molecules in a diluted gas. From this
distribution function he got a time decreasing
function of time, the so-called ${\cal H}$-function, which he identified 
with minus the entropy, obtaining then a solution for the irreversibility problem at
molecular level!

The persistent objections of influential opponents
such as Ernest Mach (1838 - 1916) and Friedrich Ostwald (1853 -
1932), led Boltzmann into depression and eventually to his suicide in 5th September 1906 at
the age of 62 at Duino, the 
seaside resort on the Adriatico coast nearby Trieste.

After Boltzmann's work, the irreversibility problem was revisited in 1907 
by Paul Ehrenfest
(1880 - 1933) and his wife, Tatyana Afanasyeva (1876 - 1964), who further
developed Boltzmann's idea of averaging over a certain region of the phase space and 
showed that the averaged ${\cal H}$-function would remain strictly decreasing in the
thermodynamical limit.

In 1928, Wolfgang Pauli (1869 - 1958) when considering the problem of
transitions in the context of quantum mechanical perturbation
theory, showed that satisfying the Second Principle of
Thermodynamics would require a {\it master equation}, which confirmed the 
conviction of Maxwell and Boltzmann 
that irreversibility should emerge from the procedure of inferring macroscopic behaviour out of microscopic physics.

More recently, Prigogine (1917 - 2003) and collaborators considered the 
radical proposal that 
irreversible behaviour should be already incorporated at the
microscopic level (see e.g. Ref. \cite{Prigogine2} for a general
discussion). However, this suggestion is not universally accepted. 

Of course, besides understanding how the irreversible behaviour of macroscopic
systems did arise from microphysics, physicists identified several phenomena whose
behaviour exhibit an immutable flow from past to present, from
present to future. The British astrophysicist and cosmologist Arthur Eddington (1882 -
1944) \cite{Eddington}, coined the term ``arrow of time'' to refer to this
evolutionary behaviour (see also Refs. \cite{Davies,Zeh} for an updated discussion). 
Let us briefly describe and enumerate these phenomena:

\vspace{0.3cm}

\noindent 1) The direction of the growth of the entropy in irreversible and dissipative macroscopic phenomena 
coincides with the flow of time;

\vspace{0.3cm}

\noindent 2) The propagation of electromagnetic
radiation, converging from the infinite to a source is not observed,
even though this being a possible solution of the Maxwell's equations of the electromagnetic field;

\vspace{0.3cm}

\noindent 3) The  irreversible collapse of the wave function of a quantum system after the measurement process; 

\vspace{0.3cm}

\noindent 4) The exponential degradation in time of systems and
the exponential growth of self-organised systems (for a
sufficiently abundant supply of resources). The development of
complex systems has led some authors to
refer to this behaviour as ``creative
evolution'', ``arrow of life'' or ``physics of becoming''
\cite{Prigogine1,Coveney,Prigogine2,BernardiniOB2022a}; 

\vspace{0.3cm}

\noindent 5) The discovery of the CP-symmetry violation in the
$K^0 - \bar{K}^0$ system due to weak interactions, allows for inferring from the CPT-theorem, a
fundamental cornerstone of quantum field theory, that the
T-symmetry is also violated\footnote{C, denotes charge conjugation, that is, particle-antiparticle operation; P, denotes parity or mirror 
image operation; T, means time inversion operation. The CPT-theorem establishes that the result of these three operations together leave states unchanged in local quantum field theories.} 
This means that there exists, at a fundamental level, an intrinsic irreversibility. 

\noindent
It is relevant to point out that the breaking of
the CP-symmetry and the violation of the baryon number\footnote{The number that counts the content of quarks and anti-quarks of a given hadronic particle} in an expanding Universe
are conditions from which the observed baryon asymmetry of the
universe (BAU) can be set up (see, for instance, Ref. \cite{Buchmuller}). An alternative route to explain the BAU is
through the violation of the CPT-symmetry itself. This is possible
in the context of string theory (see Ref. {\cite{Bertolami97} and
references therein);

\vspace{0.3cm}

\noindent 6) Psychological time is obviously irreversible and
historical. The past can be recognised, while the future is unknown.
Presumably this can be related with the issue of
causation. A fascinating question is how this 
fundamental feature of Nature has shaped, through 
Evolution, the structure of our brain which is known to have a common cortical metrics for time, space and quantity
\cite{Walsh03};

\vspace{0.3cm}

\noindent 7) Systems bound gravitationally exhibit the so-called
gravito-thermal catastrophic behaviour \cite{Lynden-Bell}, that is, 
their entropy grows as they contract, which in turn implies that
their specific heat is negative. This feature is also shared by {\it black holes}, whose entropy is proportional to their horizon's area. This feature raises the extremely relevant 
question of determining whether the gravitational field should have an intrinsic entropy \cite{Penrose}.  
In fact, it has been recently proposed that an entropy should be assigned even to the vacuum \cite{Bertolami2021}. 

On the largest scales, the expansion of the
Universe, which is itself adiabatic, is a unique phenomenon, which was conjectured 
to be the master arrow of time from which all
others should be subordinated via the putative entropy of the gravitational field \cite{Penrose}.

\section{Time in Relativity}\label{Sec:III}

The conceptual standing of time in Relativity implies, at first, that it has 
the same status as space, even though their specific properties are somewhat different. 
The main consequence of Relativity is that it endows space-time with plastic features.  
In Special Relativity, it follows from its fundamental tenets that time in a moving frame of reference stretches   
with respect to the time in a frame of reference at rest. 
This means that clocks in motion slow down with respect to clocks at rest. Under the same conditions, space contracts. 
These are purely kinematical effects and follow from the fundamental requirement that the laws of physics are 
invariant, or are said to be covariant, in whatever frame of reference that move with respect to each other at constant velocity. 
These frames of reference are called inertial frames. 

In the General Theory of Relativity, the covariance of the laws of physics are generalised to frames of 
reference that are accelerated 
with respect to each other, that is, the General Theory concerns non-inertial frames of reference. Einstein first showed 
that these frames are {\it equivalent} to a gravitational field and this discovery allowed him to turn the General Theory 
into a theory of gravity. Furthermore, after reasoning that a light ray in an accelerated frame of reference would be seen 
by an observer at rest as following a curved trajectory, he concluded that the same would happen 
in the presence of a gravitational field. He went on to conclude that  matter-energy curves space-time. It follows that in General Relativity 
the time flow is affected by gravitational fields. 

Special and General Relativity are extremely successful
theories from the experimental point of view. General Relativity,
for instance, is very well established at Solar System level \cite{Will,BPT,OB_Paramos} and its predictions 
are multifold, ranging from the existence of
black holes and gravitational radiation to cosmological descriptions, such as the Hot Big-Bang model, 
in the context of which space-time and matter were created about 13.7 thousand millions of years ago (for extensive discussions, see, 
for instance, Refs. \cite{Weinberg,HawkingEllis,Wald}).

\subsection{Closed time-like curves and the ensued menace of causality violation}

As seen above, General Relativity admits a quite rich lore of conceptual and 
mathematical solutions. Among them one finds non-trivial geometries where time
folds into itself and gives origin closed time-like curves (CTCs). These solutions potentially violate
causality as they allow for an observer who travels along this type of curve to
return to an event that coincides with the departure event giving origin to time travel paradoxes
\cite{Thorne, Nahin}. Obviously, the arrow of time
leads to a forward local flow of time, but globally an observer can, when travelling along a CTC, return to an event in the past. 

The existence of CTCs imply paradoxes like the 
observer that kills an ancestor making his/her very existence impossible as well as the 
possibility of the flow of information from the future or from the present to the past. These violations of causality 
have also quite clear implications, for instance, to the issue of free-will.

A great number of solutions of Einstein's field equations
contain CTCS, but there are two quite conspicuous types of solutions \cite{Lobo:2002rp}. 
Solutions with a tipping over of the light cones due to a rotation about a
cylindrically symmetric axis \cite{Godel,Tipler1974,Gott,Deser1,Deser2}; and solutions that
violate the energy conditions of General Relativity.
The latter admit matter-energy distributions where local observers 
effectively measure negative energy densities. Although classical forms of matter 
do not allow for this behaviour, it is argued that quantum
fields may admit this possibility. In fact, this condition can be relaxed in alternative theories of gravity 
(see, for instance,  Refs. \cite{Lobo2011,OB_Ferreira2012}).  

Furthermore one could admit the reverse procedure and consider geometries which could lead, 
after some simple manipulations, to CTCs \cite{MT,Everett,ER,frolovnovikovTM}. 
These include the so-called traversable
wormholes \cite{Morris,Visser}, the warp drive
\cite{Alcubierre,Lobo:2004wq,Lobo:2002zf}, and the Krasnikov tube
\cite{Krasnikov}. 

Let us briefly discuss the case 
{\it traversable wormholes}. A wormhole is essentially
constituted by two mouths at different
regions of spacetime connected by
a tunnel or handle. Several types of manipulations can give origin to CTCs. 
The simplest one involves a single
wormhole mouth, which is moving with respect to the other 
mouth so to create a time shift (cf. Special Relativity) 
without affecting the internal geometry of the wormhole \cite{MT}. 
More elaborate variations of this procedure can be envisaged 
\cite{Earman}.

\subsection{Defusing the causality paradoxes}

The existence of CTCs can potentially undo irreversible events at the 
expense of creating somewhat embarrassing violations of causality. If one keeps the metric 
nature of General Relativity, hence the only way to avoid these violations is through a veto to CTCs. 
These curves can be rejected on logical or physical grounds. The latter are obviously preferable.  
The following solutions have been proposed: 

\noindent
i) In Novikov's {\it Principle of Self-Consistency} \cite{Novikov-CTCWH}, CTCs are admitted if self-consistent, 
that is, the usual causal structure might be broken, but the future can only influence without changing events in the past.
This principle implies that solutions of the laws of
physics that are locally admissible are those that are also globally
self-consistent.

\noindent
ii) Hawking's {\it Chronology Protection Conjecture} is based on  
the experimental evidence that ``we have not been invaded by hordes of tourists
from the future'' \cite{Hawk2}. The physical support for the conjecture lies on 
the observation that a relevant physical quantity, the renormalised quantum expectation of the 
stress-energy tensor diverges close to the CTC's. It is argued that the conjecture can be shown in the context 
of the quantum gravity theory, the theory that will presumably reconcile General Relativity to Quantum Mechanics. The typical scale of energy 
of this conjectured theory is about $10^{19}~GeV$, about fifteen orders of magnitude greater than the energy of the LHC collider at CERN in Geneva, Switzerland, 
concentrated at a extremely small region of space, about $10^{-35}~m$.   

\noindent
iii) It has been pointed out the CTCs arise in the limit of theories where sensible physical properties of matter break down \cite{Deser2,OB2013}. 
The generality of this situation for most of the solutions would render CTCs unphysical. 

Thus, it is clear that the existence of CTCs is still the subject of discussion and the final answer cannot be known before a deeper understanding of quantum gravity. 
In fact we shall see that the matter of time travelling becomes particularly pressing when considering the idea of a Block Universe.   

\section{Some Relevant Open Questions}\label{Sec:IV}

Let us briefly discuss here some ideas about the nature of
space-time that are still a matter of lively debate: 

{\it Time crystals}

Time crystals are quantum mechanical systems in their lowest energy state that exhibit a periodic motion \cite{Wilczek1,Wilczek2}, 
analogously to usual crystals that are a periodic repetition of atoms in space. 
Their existence are associated to a putative spontaneous breaking of discrete time translational symmetry, 
which is thought to be impossible in Quantum Mechanics. However, 
it is believed that evidence of time crystals have been found experimentally in many systems (see Ref. \cite{Sacha} for a review). 
A recent proposal suggests that time crystals are an emergent feature 
of phase-space noncommutative quantum mechanics \cite{BernardiniOB2022b}.

{\it Block Universe}

It is assumed that as time evolves, gone events no longer exist, meaning 
that as the future is not there yet, only the ``present" is actual. 
An interesting alternative description of space-time is the so-called 
Block Universe, in which a preferred ``present'' is non-existent and
past is equally present. Thus, all points in time are
equally valid frames of reference and a specific instant that
lies in the past or in the future is frame dependent. In principle, there is no logical 
contradiction in a space-time block like that, however even though observers can 
experience the subjective flow of time, an universal present is ruled out by Special Relativity 
where no simultaneity exists. Common sense and irreversible phenomena seem to contradict 
the possibility of a Block Universe. Indeed, we refer to Ref. \cite{Ellis:2006sq} a consistent discussion  
on the objections to the Block Universe viewpoint.

It is evident that in a Block Universe, the causality paradoxes are particularly pressing and demand for a 
categoric answer about the existence of CTCs.

{\it Specious Present}

The so-called ``specious present", has been the subject of discussion in experimental psychology and philosophy for quite some time. Our perception of the present 
is such that is difficult, if not impossible, to precise the span of time it comprises. 
It is a well established fact that in our experience, objects are given to us as being of the present, 
but the part of time referred to by the data is a very different one in what it concerns past and future.
The specious present has also been encountered in a completely independently way when dealing with the engineering 
problem of transmission of visual and audio information in a synchronised way. 

In fact, experiments show that our brain copes well when a sensory delay between visual and audio information is smaller than $125~ms$ 
for present-past transmission, but is much shorter, about $45~ms$, for present-future transmission (see, for instance, Ref. \cite{Stanford} for a discussion). 
How to interpret this time asymmetry? Does it reflect the fundamental causality discussed by the philosopher David Hume? 
Is it a consequence of Evolution, which causally wired our brain to cope with sensory data?

\section{Conclusions}\label{Sec:Conclusion}

In this brief report we attempted to discuss in a self consistent way the nature of time and the most 
conspicuous implications related with various conceptions of this hidden and mysterious ingredient of the physical description. 
After a short discussion of some specific philosophical ideas about time, we spelled out how the description of macroscopic phenomena 
through the Second Law of Thermodynamics evolved so to allow us to understanding how irreversibility arises from 
microphysics, despite the symmetry of the fundamental evolution equations, classical and
quantum, under time reversal. Indeed, in the context of
Statistical Mechanics\footnote{The branch of physics that describes microphysics and its matching to the macroscopic description achieved through the Laws of Thermodynamics.}, 
it is assumed that the absolute deterministic behaviour is lost as the macroscopic
description necessarily averages out the micro-properties of the
systems. 

We have then discussed how in the Relativity revolution the physical 
description suggested a merged space-time depiction of the events so to endow space-time 
with plastic properties. 
We also elucidated how it followed  
that time flows at different rates for observers in different frames of reference. As seen, this is radically different from 
the description according Newtonian physics, where time was assumed to flow at a constant rate
for all observers. In the context of General Relativity, space-time is curved by matter-energy and 
the flow of time is dependent of the strength of the gravitational field. The space-time curvature leads 
to completely new phenomena such as black holes, gravitational waves and a great variety of cosmological 
spaces, including the one with a Big Bang, which about 
13.7 thousand millions years ago, gave origin to matter and space-time itself. 

General Relativity also admits solutions of its field equations in which time folds into itself, giving origin to CTCs. This surprising possibility can lead, as 
discussed, to causality paradoxes. We have then presented some putative solutions to these embarrassing paradoxes. Even though, these solutions seem quite sensible 
as they clearly do the job of protecting causality, presumably, a definite answer about the existence of CTCs requires an encompassing theory of quantum gravity.  

Finally, we have closed our report with some brief discussions about topics that are intimately related with the nature of time and how it is perceived, 
namely, time crystals, the Block Universe proposal 
and the question about how our brain processes sensorial information, the specious present question, and its well established temporal asymmetry. 

As a closing remark we could say that any attempt to encompass all aspects of the problem of time is bound to fail. This author was well aware of this essential limitation 
and pursued no more than to present, through some broad strokes, a brief account on how physicists approach the fundamental problem of time. 
The concept of time is at the very heart of the epistemological questions associated with the physical description of phenomena, 
both at macroscopic and at microscopic scales. However, this is necessarily a partial view of the matter. The ubiquity of time 
in any form of articulate thinking puts it conceptually at the very heart of any discussion about consciousness and 
its emergence at biological, physical and philosophical levels. Time, being the History of everything is the horizon of all forms of thinking. 
The sought for an explanation about the most essential nature of time can only lead us to the source of new questions.


\begin{thebibliography}{99}


\bibitem{Farrell2005} J. Farrell, {\it The Day Without Yesterday: Lema\^itre, Einstein, and the Birth of Modern Cosmology}
(Basic Books, New York 2005).

\bibitem{Bertolami2006} O. Bertolami, {\it O Livro das Escolhas C\'osmicas}
(Gradiva, Lisboa 2006).


\bibitem{Bertolami2007} O. Bertolami, ``The adventures of Spacetime'' 
in {\it Relativity and the Dimensionality of the World},
Ed. V. Petkov, Springer Fundamental Theories of Physics 153
(Springer, AA Dordrecht The Netherlands 2007).


\bibitem{Minkowski} H. Minkowski, ``Raum und Zeit'', 1908.
Also in A.H. Lorentz, A. Einstein e H. Minkowski,
{\it O Princ\'\i pio da Relatividade}
(Funda\c c\~ao Calouste Gulbenkian, Lisboa 1978).

\bibitem{Bertolami2008a}
O.~Bertolami,
``The mystical formula and the mystery of Khronos'' in 
{\it Minkowski Spacetime: A Hundred Years Later}, Ed. V. Petkov, 
Springer Fundamental Theories of Physics (2010).

\bibitem{Bertolami_Lobo} O.~Bertolami and F. Lobo, `Time and Causation", 
NeuroQuantol. 7, 1-15 (2009); arXiv:0902.0559 [gr-qc].  

\bibitem{Russell} B. Russell, {\it History of Western Philosophy}
(Counterpoint, London 1946).

\bibitem{Prigogine1} I. Progine and I Stengers, {\it La Nouvelle Alliance}
(Gallimard, Paris 1979).

\bibitem{Montheron} {\it Des hommes de science aux prises avec le temps},  
Group de Matheron (Presse Polytechniques e Universitaires
Romandes, Laussane 1992).

\bibitem{Ellis} G.F.R. Ellis, ``Physics in a Real Universe:
Time and Space-Time'' in {\it Relativity and the Dimensionality of the World},
Ed. V. Petkov, Springer Fundamental Theories of Physics 153
(Springer, AA Dordrecht The Netherlands 2007).

\bibitem{Tipler80} F.J. Tipler, Essays in General Relativity,
Festschrift for A. Taub, Ed. F.J. Tipler (Academic Press, 1980).

\bibitem{Coveney} P. Coveney and R. Highfield, {\it The Arrow of Time}
(Fawcett Columbine, New York 1990).

\bibitem{Seiberg} N. Seiberg, ``Emergent Spacetime", arXiv: hep-th/0601234.

\bibitem{Newton}
I. Newton, (1726), {\it The Principia}, 3rd edition. Translated by
I. Bernard Cohen and Anne Whitman, (University of California
Press, Berkeley, 1999).

\bibitem{Canales} Jimena Canales, {\it The Physicist and the Philosopher: Einstein, Bergson, and the 
Debate That Changed Our Understanding of Time} (Princeton University Press 2015). 

\bibitem{Eddington} A. Eddington, {\it The Nature of the Physical World}
(Cambridge University Press, Cambridge 1928).

\bibitem{Davies} P.C.W. Davies, {\it The Physics of the Time Asymmetry}
(California University Press, Berkeley 1974).

\bibitem{Zeh} H.D. Zeh, ``The Physical Foundation of the Direction of Time'',
Heidelberg University Preprint 1988.

\bibitem{Prigogine2} I. Prigogine, {\it From Being to Becoming} 
(W.H. Freeman \& Co. 1980).

\bibitem{BernardiniOB2022a} A. Bernardini and O. Bertolami, `Non-commutative phase-space Lotka-Volterra dynamics: the quantum analogue",  Phys. Rev. {E 106}, 024202 (2022). . 

\bibitem{Buchmuller} W. Buchm\"uller, ``Baryogenesis - 40 Years later'',
arXiv: 0710.5857[hep-ph].

\bibitem{Bertolami97} O. Bertolami, D. Colladay, V.A. Kosteleck\'y and R.
Potting, ``CPT violation and baryogenesis'', Phys. Lett. B. {\bf 395}, 
178 (1997).

\bibitem{Walsh03} V. Walsh, ``A Theory of magnitude: common cortical 
metrics of time, space and quantity'', 
Trends in Cognitive Sciences, {\bf 7}, 483 (2003).

\bibitem{Lynden-Bell} D. Lynden-Bell, 
Mont. Not. Roy. Astr. Soc. {\bf 123}, 447 (1962).

\bibitem{Penrose} R. Penrose, ``Singularities and time-asymmetry", in {\it General Relativity: An Einstein Centenary Survey}, Eds. S.
Hawking and W. Israel (Cambridge University Press 1979).

\bibitem{Bertolami2021} O. Bertolami,  ``Inflation, phase transitions and the cosmological constant", Gen. Rel. Grav.  {\bf 53},  
11, 1006 (2021).

\bibitem{Will} C. Will, ``The Confrontation between General Relativity
and Experiment'', arXiv: gr-qc/0510072.

\bibitem{BPT} O. Bertolami, J. P\'aramos and S. Turyshev,
``General Theory of Relativity: Will it survive the next decade?'' in
Lasers, Clocks, and Drag-Free: Technologies for Future Exploration in
Space and Tests of Gravity: Proceedings. Edited by H. Dittus, C. Laemmerzahl,
S. Turyshev. Springer Verlag, 2006; arxiv: gr-qc/0602016.

\bibitem{OB_Paramos} O. Bertolami and J. P\'aramos, ``The experimental status of Special and General Relativity", 
in {\it Springer Handbook of Spacetime}, 463-483 (2014). 

\bibitem{Weinberg} S. Weinberg, {\it Gravitation and Cosmology Principles and Applications of the General Theory of Relativity} (Wiley 1972). 

\bibitem{HawkingEllis}
S. W. Hawking and G. F. R. Ellis, {\it The Large Scale Structure
of Spacetime} (Cambridge University Press, Cambridge 1973).

\bibitem{Wald}
R. M. Wald, {\it General Relativity} (University of Chicago Press,
Chicago 1984).

\bibitem{Thorne} K. S. Thorne, ``Closed Timelike Curves'', General Relativity and Gravitation, Eds. J.L. Gleiser 
et al. (Institute of Physics, Bristol 1992).

\bibitem{Nahin}
P. J. Nahin, {\it Time Machines: Time Travel in Physics,
Metaphysics and Science Fiction} (Springer-Verlag and AIP Press,
New York 1999).

\bibitem{Visser}
M. Visser, {\it Lorentzian Wormholes: From Einstein to Hawking} 
(American Institute of Physics, New York 1995).

\bibitem{Lobo:2002rp}
F.~Lobo and P.~Crawford,
``Time, closed timelike curves and causality'', 
NATO Sci.\ Ser.\ II {\bf 95}, 289 (2003); arXiv: gr-qc/0206078.
  
\bibitem{Godel}
K. G\"{o}del, ``An Example of a New Type of Cosmological Solution
of Einstein's Field Equations of Gravitation'', Rev. Mod. Phys.
{\bf 21}, 447 (1949).

\bibitem{Tipler1974}
F. J. Tipler, ``Rotating Cylinders and the Possibility of Global
Causality Violation'', Phys. Rev. D {\bf 9}, 2203 (1974);

\bibitem{Gott}
J. R. Gott, ``Closed Timelike Curves Produced by Pairs of Moving
Cosmic Strings: Exact Solutions'', Phys. Rev. Lett. {\bf 66}, 1126
(1991).

\bibitem{Deser1}
S. Deser, R. Jackiw and G. t'Hooft, ``Physical cosmic strings do
not generate closed timelike curves'', Phys. Rev. Lett. {\bf 68}
267 (1992).

\bibitem{Deser2}
S. Deser, ``Physical obstacles to time-travel'', Class. Quant.
Grav. {\bf 10}, S67 (1993).

\bibitem{Lobo2011} N.M. Garcia, F. Lobo, ``Nonminimal curvature-matter coupled wormholes with matter satisfying the null energy condition", 
Class. Quant. Grav. {\bf 28}, 085018  (2011).

\bibitem{OB_Ferreira2012} O. Bertolami, R.Z. Ferreira, ``Traversable Wormholes and Time Machines in non-minimally coupled curvature-matter 
f(R) theories", Phys. Rev. D {\bf 85}, 104050 (2012).

\bibitem{MT}
M. Morris, K. S. Thorne and U. Yurtsever, ``Wormholes, Time
Machines and the Weak Energy Condition'', Phys. Rev. Lett. {\bf
61}, 1446 (1988).

\bibitem{Everett}
A. E. Everett, ``Warp Drive and Causality'', Phys. Rev. D {\bf
53}, 7365 (1996).

\bibitem{ER}
A. E. Everett and T. A. Roman, ``A Superluminal Subway: The
Krasnikov Tube'', Phys. Rev. D {\bf 56}, 2100 (1997)

\bibitem{frolovnovikovTM}
V. P. Frolov and I. D. Novikov, ``Physical effects in wormholes
and time machines'', Phys. Rev. D {\bf 42}, 1057 (1990).

\bibitem{Morris}
M. Morris and K. S. Thorne, ``Wormholes in Spacetime and Their Use
for Interstellar Travel: A Tool for Teaching General Relativity'', 
Am. Journ. Phys.  {\bf 56}, 395 (1998).

\bibitem{Alcubierre}
M. Alcubierre, ``The Warp Drive: Hyper-Fast Travel within General
Relativity'', Class. Quant. Grav. {\bf 11}, L73 (1994).

\bibitem{Lobo:2004wq}
F.~S.~N.~Lobo and M.~Visser, 
``Fundamental limitations on `warp drive' spacetimes'',  
Class.\ Quant.\ Grav.\  {\bf 21}, 5871 (2004).

\bibitem{Lobo:2002zf}
F.~Lobo and P.~Crawford,
``Weak energy condition violation and superluminal travel'',  
Lect.\ Notes Phys.\  {\bf 617}, 277 (2003); arXiv: gr-qc/0204038.
  
\bibitem{Krasnikov}
S. V. Krasnikov, ``Hyper-Fast Interstellar Travel in General
Relativity'', Phys. Rev. D {\bf 57}, 4760 (1998).

\bibitem{Earman}
J. Earman, {\it Bangs, Crunches, Whimpers, and Shrieks:
Singularities and Acausalities in Relativistic Spacetimes} (Oxford
University Press, Oxford 1995).

\bibitem{Novikov-CTCWH}
I. D. Novikov, ``An analysis of the operation of a time machine'', 
Sov. Phys. JETP {\bf 68} 3 (1989).

\bibitem{Hawk2}
S. W. Hawking, ``Chronology Protection Conjecture'', Phys. Rev. D
{\bf 56}, 4745 (1992).

\bibitem{OB2013} O. Bertolami, ``Zen and the Art of Space-Time Manufacturing", 
EPJ Web Conf. {\bf 58}, 02001 (2013). 

\bibitem{Wilczek1} F. Wilczek, ``Quantum Time Crystals", Phys. Rev. Lett. {\bf109}, 160401 (2012).

\bibitem{Wilczek2} A. Shapere and F. Wilczek, ``Classical Time Crystals". Phys. Rev. Lett. {\bf 109}, 160402 (2012).

\bibitem{Sacha}  K. Sacha and J. Zakrzewski, ``Time crystals: a review", Rep. Prog. Phys. {\bf 81}, 016401 (2018).

\bibitem{BernardiniOB2022b} A. Bernardini and O. Bertolami,  ``Emergent time crystals from phase-space noncommutative quantum mechanics", Phys. Lett. {\bf 835}, 137549 (2022).  

\bibitem{Ellis:2006sq}
G.~F.~R.~Ellis,
``Physics in the real universe: Time and spacetime'', 
Gen.\ Rel.\ Grav.\  {\bf 38}, 1797 (2006).

\bibitem{Stanford} The Specious Present: Further Issues, Stanford Encyclopedia of Philosophy (https://plato.stanford.edu/entries/consciousness-temporal/specious-present.html). 





\end{thebibliography}
\end{document}